\documentclass[a4paper]{article}

\usepackage{18lomcon}        
\usepackage{cite}            
\usepackage{epsfig}          
\usepackage{amsmath,amssymb}

\newcommand{\ind}[2]{^{#1}_{\text{#2}}}
\def\epem{e^{+}e^{-}}
\def\nf{n_{f}}

\begin{document}

\title{CALCULATION OF THE $R$--RATIO OF {\large$\epem\!\to\,$}HADRONS \\[1mm]
AT THE HIGHER--LOOP LEVELS}

\author{A.V.~Nesterenko \email{nesterav@theor.jinr.ru,~nesterav@gmail.com}}

\affiliation{BLTPh JINR, Dubna, 141980, Russian Federation}

\maketitle

\begin{abstract}
The calculation of the $R$--ratio of electron--positron annihilation into
hadrons is discussed. The method, which enables one to properly account
for all the effects due to continuation of the spacelike perturbative
results into the timelike domain at an arbitrary loop level, is
delineated.
\end{abstract}

The theoretical description of a variety of the strong interaction
processes is inherently based on the hadronic vacuum polarization
function~$\Pi(q^2)$, which is defined as the scalar part of the hadronic
vacuum polarization tensor
\begin{equation}
\label{P_Def}
\Pi_{\mu\nu}(q^2) = i \!\int\! d^4 x\,e^{i q x} \bigl\langle 0 \bigl|\,
T\left\{J_{\mu}(x)\, J_{\nu}(0)\right\} \bigr| 0 \bigr\rangle =
\frac{i}{12\pi^2} (q_{\mu}q_{\nu} - g_{\mu\nu}q^2) \Pi(q^2),
\end{equation}
the related Adler function~\cite{Adler}
\begin{equation}
\label{Adler_Def}
D(Q^2) = - \frac{d\, \Pi(-Q^2)}{d \ln Q^2},
\end{equation}
and the function~$R(s)$
\begin{equation}
\label{R_Def}
R(s) = \frac{1}{2 \pi i} \lim_{\varepsilon \to 0_{+}}
\Bigl[\Pi(s + i \varepsilon) - \Pi(s - i \varepsilon)\Bigr] =
\frac{\sigma(\epem \to \text{hadrons}; s)}
{\sigma(\epem \to \mu^{+}\mu^{-}; s)},
\end{equation}
which is identified with the so--called $R$--ratio of electron--positron
annihilation into hadrons. The functions~$\Pi(q^2)$~(\ref{P_Def})
and~$D(Q^2)$~(\ref{Adler_Def}), being the functions of the spacelike
kinematic variable~$Q^2 = -q^2 > 0$, can be directly accessed within
QCD~perturbation theory, whereas the $R$--ratio~(\ref{R_Def}), being the
function of the timelike kinematic variable~$s = q^2 > 0$, can be
described only by making use of the relevant dispersion relations.
Specifically, the relation, which expresses the $R$--ratio~(\ref{R_Def})
in terms of the theoretically calculable Adler function and provides a
native way to properly account for the effects due to continuation of the
spacelike perturbative results into the timelike domain, can be obtained
by integrating Eq.~(\ref{Adler_Def}) in finite limits, that
yields~\cite{Rad82, KP82}
\begin{equation}
\label{R_Disp2}
R(s) =  \frac{1}{2 \pi i} \lim_{\varepsilon \to 0_{+}}
\int_{s + i \varepsilon}^{s - i \varepsilon}
D(-\zeta)\,\frac{d \zeta}{\zeta}.
\end{equation}
The integration contour in this equation lies in the region of analyticity
of the integrand in the complex $\zeta$--plane.

It is necessary to outline that the dispersion relations, which express
the functions~$\Pi(q^2)$, $R(s)$, and~$D(Q^2)$ in terms of each other,
rely only on the kinematics of the process on hand and involve neither
model--dependent phenomenological assumptions nor additional
approximations. In~turn, such relations impose a number of strict physical
intrinsically nonperturbative constraints on the functions~$\Pi(q^2)$,
$R(s)$, and~$D(Q^2)$, that should certainly be accounted for when one
comes out of the limits of applicability of the QCD perturbation theory.
It is worthwhile to note also that these nonperturbative restrictions have
been merged with corresponding perturbative input in the framework of
dispersively improved perturbation theory~(DPT)~\cite{Book, DPT1, DPT2}
(its preliminary formulation was discussed in Ref.~\cite{DPTPrelim}). In
particular, the DPT enables one to overcome some inherent difficulties of
the QCD perturbation theory and extend its applicability range towards the
infrared domain, see book~\cite{Book} and references therein.

This study is primarily focused on the theoretical description of the
$R$--ratio of electron--positron annihilation into hadrons~(\ref{R_Def})
at moderate and high energies, so that the nonperturbative aspects of the
strong interactions will be disregarded hereinafter. For this purpose the
effects due to the masses of the involved particles can be safely
neglected (the impact of such effects on the low--energy behavior of the
functions on hand was discussed in, e.g., Refs.~\cite{Book, DPT1, DPT2,
DPT3}). In the massless limit the relation~(\ref{R_Disp2}) can be
represented as
\begin{equation}
\label{RProp}
R^{(\ell)}(s) = 1 + r^{(\ell)}(s),
\quad
r^{(\ell)}(s) =
\int_{s}^{\infty}\rho^{(\ell)}(\sigma)\,
\frac{d \sigma}{\sigma},
\end{equation}
where
\begin{equation}
\label{RhoDef}
\rho^{(\ell)}(\sigma) =
\frac{1}{2 \pi i} \lim_{\varepsilon \to 0_{+}}
\Bigl[d^{(\ell)}(-\sigma - i \varepsilon) -
d^{(\ell)}(-\sigma + i \varepsilon)\Bigr]
\end{equation}
stands for the spectral function and~$d^{(\ell)}(Q^2)$ denotes the
$\ell$--loop strong correction to the Adler function~(\ref{Adler_Def}).
As~mentioned above, only perturbative contributions will be retained in
Eq.~(\ref{RhoDef}) in what follows, that makes Eq.~(\ref{RProp}) identical
to that of the ``Analytic perturbation theory''~\cite{APT0} (for some of
its applications see Refs.~\cite{APT1, APT2, APT3, APT4, APT5, APT6,
APT7a, APT7b, APT8a, APT8b, APT9, APT10, 12dAnQCD}). A discussion of the
nonperturbative terms in the spectral density~$\rho^{(\ell)}(\sigma)$ can
be found in, e.g., Refs.~\cite{PRD6264, Review, MPLA1516, APTCSB}.

The perturbative expression for the Adler function~(\ref{Adler_Def}) takes
the form of the power series in the so--called QCD couplant
$a\ind{(\ell)}{s}(Q^2) = \alpha\ind{(\ell)}{s}(Q^2)\, \beta_{0}/(4\pi)$
\begin{equation}
\label{AdlerPert}
D^{(\ell)}_{{\rm pert}}(Q^2) = 1 + d^{(\ell)}_{{\rm pert}}(Q^2),
\quad
d^{(\ell)}_{{\rm pert}}(Q^2) = \sum_{j=1}^{\ell} d_{j}
\left[a\ind{(\ell)}{s}(Q^2)\right]^{j}.
\end{equation}
Here~$\ell$ specifies the loop level, $d_1 = 4/\beta_0$, $\beta_0 = 11 -
2\nf/3$, $\nf$~is the number of active flavors, the common prefactor
$N_{\text{c}}\sum_{f=1}^{\nf} Q_{f}^{2}$ is omitted throughout,
$N_{\text{c}}=3$ denotes the number of colors, and $Q_{f}$~stands for the
electric charge of $f$--th quark. The QCD couplant entering
Eq.~(\ref{AdlerPert}) can be represented~as
\begin{equation}
\label{AItGen}
a\ind{(\ell)}{s}(Q^2) =
\sum_{n=1}^{\ell}\sum_{m=0}^{n-1} b^{m}_{n}\,
\frac{\ln^{m}(\ln z)}{\ln^n z},
\end{equation}
where $z=Q^2/\Lambda^2$, $b^{m}_{n}$ is the combination of the
$\beta$~function perturbative expansion coefficients ($b^{0}_{1}=1$,
$b^{0}_{2}=0$, $b^{1}_{2}=-\beta_{1}/\beta^{2}_{0}$,~etc.), and~$\Lambda$
denotes the QCD scale parameter. The Adler function perturbative expansion
coefficients~$d_{j}$ were calculated up to the four--loop
level~\cite{RPert4L}, whereas the $\beta$~function perturbative expansion
coefficients~$\beta_{j}$ are available up to the five--loop
level~\cite{Beta5L}.

Since the calculation of the spectral
function~$\rho^{(\ell)}(\sigma)$~(\ref{RhoDef}) becomes rather cumbrous
beyond the one--loop level (the explicit expressions
for~$\rho^{(\ell)}(\sigma)$ at first four loop levels can be found in
Ref.~\cite{CPC}), one commonly re--expands the strong
correction~$r^{(\ell)}(s)$~(\ref{RProp}) at high energies, that eventually
leads to~\cite{Pi2termsHO, RpertHO}
\begin{align}
\label{Pi2TermsGen}
r^{(\ell)}(s) & =
\sum_{j=1}^{\ell} d_{j} \left[a^{(\ell)}_{{\rm s}}(|s|)\right]^{j} -
\sum_{j=1}^{\ell} d_{j}
\sum_{n=1}^{\infty}\! \frac{(-1)^{n+1}}{(2n+1)!}\pi^{2n} \times
\nonumber \\ & \hspace{-7mm} \times
\!\sum_{k_{1}=0}^{\ell-1}
\!\ldots\!
\sum_{k_{2n}=0}^{\ell-1}\!
\left(\prod_{p=1}^{2n}B_{k_{p}}\!\right)\!
\left[\prod_{t=0}^{2n-1}\!
\Bigl(j+t+k_{1}+k_{2}+\ldots+k_{t}\Bigr)\!\right]\! \times
\nonumber \\ & \hspace{-7mm} \times\!
\Bigl[a^{(\ell)}_{{\rm s}}(|s|)\Bigr]^{j+2n+k_{1}+k_{2}+\ldots+k_{2n}},
\qquad \sqrt{s}/\Lambda>\exp(\pi/2).
\end{align}
It is necessary to emphasize that the re--expansion~(\ref{Pi2TermsGen}) is
valid only for~$\sqrt{s}/\Lambda > \exp(\pi/2) \simeq 4.81$, and it
converges rather slowly when the energy scale approaches this value. If
the number of terms retained on the right--hand side of
Eq.~(\ref{Pi2TermsGen}) is large enough, then it can provide quite
accurate approximation of the strong correction to
the~$R$--ratio~(\ref{RProp}). However, one usually truncates the
re--expansion~(\ref{Pi2TermsGen}) at the order~$\ell$, thereby neglecting
all the higher--order $\pi^2$--terms (though, the latter may not
necessarily be negligible, see Ref.~\cite{RpertHO}), that results in the
expression commonly employed in the practical applications:
\begin{equation}
\label{RAppr}
R^{(\ell)}_{{\rm appr}}(s) = 1 + r^{(\ell)}_{{\rm appr}}(s),
\quad
r^{(\ell)}_{{\rm appr}}(s) =\!
\sum_{j=1}^{\ell} r_{j} \!\left[a^{(\ell)}_{{\rm s}}(|s|)\right]^{j}\!,
\quad
r_{j} = d_{j} - \delta_{j}.
\end{equation}
Here $d_j$ denote the Adler function perturbative expansion
coefficients~(\ref{AdlerPert}) and~$\delta_{j}$ embody the contributions
of relevant~$\pi^2$--terms~(\ref{Pi2TermsGen}), see Refs.~\cite{Pi2Terms1,
RPert5LEstim1, ProsperiAlpha, Book, Pi2termsHO, RpertHO}.

At the same time, the explicit expression for the perturbative spectral
function entering Eq.~(\ref{RProp}) can be calculated at an arbitrary loop
level (it is assumed that the involved perturbative coefficients~$d_j$
and~$\beta_j$ are known) by making use of the method developed in
Ref.~\cite{RpertHO}, namely
\begin{align}
\label{RhoPertHO}
\rho^{(\ell)}(\sigma) & =
\sum_{j=1}^{\ell} d_{j} \sum_{k=0}^{K(j)}
\binom{j}{2k+1} (-1)^{k}\, \pi^{2k} \times
\nonumber \\
& \times
\Biggl[\sum_{n=1}^{\ell}\sum_{m=0}^{n-1} b^{m}_{n}\, u_{n}^{m}(\sigma)\Biggr]^{j-2k-1}
\Biggl[\sum_{n=1}^{\ell}\sum_{m=0}^{n-1} b^{m}_{n}\, v_{n}^{m}(\sigma)\Biggr]^{2k+1}.
\end{align}
In this equation
\begin{align}
\label{Unm}
u_{n}^{m}(\sigma) & =
\begin{cases}
u_{n}^{0}(\sigma),& \text{if $\, m=0$},\\
u_{n}^{0}(\sigma) u_{0}^{m}(\sigma) -
\pi^2 v_{n}^{0}(\sigma) v_{0}^{m}(\sigma),\quad& \text{if $\, m \ge 1$},
\end{cases}
\\
\label{Vnm}
v_{n}^{m}(\sigma) & =
\begin{cases}
v_{n}^{0}(\sigma),& \text{if $\, m=0$},\\
v_{n}^{0}(\sigma) u_{0}^{m}(\sigma) +
u_{n}^{0}(\sigma) v_{0}^{m}(\sigma),\quad\quad\,& \text{if $\, m \ge 1$},
\end{cases}
\\
\label{Un0Def}
u_{n}^{0}(\sigma) & = \frac{1}{(y^2 + \pi^2)^{n}}
\sum_{k=0}^{K(n+1)} \binom{n}{2k} (-1)^{k} \pi^{2k} y^{n-2k},
\\
\label{Vn0Def}
v_{n}^{0}(\sigma) & = \frac{1}{(y^2 + \pi^2)^{n}}
\sum_{k=0}^{K(n)} \binom{n}{2k+1} (-1)^{k} \pi^{2k} y^{n-2k-1},
\\
\label{U0mDef}
u_{0}^{m}(\sigma) & = \sum\limits_{k=0}^{K(m+1)}\binom{m}{2k}
(-1)^{k}\pi^{2k} \Bigl[L_{1}(y)\Bigr]^{m-2k}\, \Bigl[L_{2}(y)\Bigr]^{2k},
\\
\label{V0mDef}
v_{0}^{m}(\sigma) & = \sum\limits_{k=0}^{K(m)}\binom{m}{2k+1}
(-1)^{k+1}\pi^{2k} \Bigl[L_{1}(y)\Bigr]^{m-2k-1}\, \Bigl[L_{2}(y)\Bigr]^{2k+1},
\end{align}
\begin{equation}
\label{L12Def}
L_{1}(y) \!=\! \ln\!\sqrt{y^{2}+\pi^{2}},
\quad
L_{2}(y) \!=\! \frac{1}{2} - \frac{1}{\pi}\arctan\!\left(\frac{y}{\pi}\right),
\end{equation}
$K(j) = [(j-2)+(j\;\text{mod}\;2)]/2$, $y=\ln(\sigma/\Lambda^2)$, and~$n
\ge 1$ is assumed. In~turn, Eq.~(\ref{RhoPertHO}) enables one to properly
account for the effects due to continuation of the spacelike perturbative
results into the timelike domain at an arbitrary loop level, that plays a
valuable role in the studies of a variety of the strong interaction
processes, see paper~\cite{RpertHO} and references therein for the details.

\end{document}